\documentclass[10pt,journal,twoside]{IEEEtran}

\usepackage{graphicx}
\usepackage{color}
\usepackage{placeins}
\usepackage{float}
\usepackage{tabularx,colortbl}
\usepackage{multirow}
\usepackage{makecell}
\usepackage{CJK}
\usepackage{amsmath,bm}
\usepackage{cite}
\usepackage{amsthm}
\usepackage{extarrows}
\usepackage{amssymb}
\usepackage[table]{xcolor}

\begin{document}
%
\title{
Auto-Weighted Layer Representation Based View Synthesis Distortion Estimation for 3-D Video Coding
}
%
%
%

\author{Jian Jin,~\IEEEmembership{Member,~IEEE}, 
Xingxing Zhang,
Lili Meng,
Weisi Lin,~\IEEEmembership{Fellow,~IEEE},

Jie Liang,~\IEEEmembership{Senior Member,~IEEE},
Huaxiang Zhang,
Yao Zhao,~\IEEEmembership{Senior Member,~IEEE}
\thanks{Copyright \copyright 20XX IEEE. Personal use of this material is permitted. However, permission to use this material for any other purposes must be obtained from the IEEE by sending an email to pubs-permissions@ieee.org. \emph{(Corresponding author: Weisi Lin.)}}
\thanks{J. Jin and W. Lin are with the School of Computer Science and Engineering, Nanyang Technological University, 639798, Singapore and also with Alibaba-NTU Singapore Joint Research Institute, Nanyang Technological University, 639798, Singapore. E-mail: jian.jin@ntu.edu.sg; wslin@ntu.edu.sg.}
\thanks{X. Zhang is with the Department of Computer Science and Technology, Tsinghua University, Beijing, 100084, China. E-mail: xxzhang2020@mail.tsinghua.edu.cn.}
\thanks{L. Meng and H. Zhang are with the School of Information Science and Engineering, Shandong Normal University, Jinan, 250014, China. E-mail: mengll\_83@hotmail.com; huaxzhang@hotmail.com.}
\thanks{L. Jie is with the School of Engineering
Science, Simon Fraser University, Canada. Email: jiel@sfu.ca.}
\thanks{Y. Zhao is with the Institute of Information Science, Beijing Jiao Tong University and also with the Beijing Key Laboratory of Advanced Information Science and Network Technology, 100044, China. E-mail: yzhao@bjtu.edu.cn.}

}

\markboth{Submitted to IEEE Transactions on Multimedia.}{Jin {\it \lowercase{et al.}}: {
Auto-Weighted Layer Representation Based View Synthesis Distortion Estimation for 3-D Video Coding
}
}
%



\maketitle

\begin{abstract}

Recently, various view synthesis distortion estimation models have been studied to better serve for 3-D video coding. However, they can hardly model the relationship quantitatively among different levels of depth changes, texture degeneration, and the view synthesis distortion (VSD), which is crucial for rate-distortion optimization and rate allocation. In this paper, an auto-weighted layer representation based view synthesis distortion estimation model is developed. Firstly, the sub-VSD (S-VSD) is defined according to the level of depth changes and their associated texture degeneration. After that, a set of theoretical derivations demonstrate that the VSD can be approximately decomposed into the S-VSDs multiplied by their associated weights. To obtain the S-VSDs, a layer-based representation of S-VSD is developed, where all the pixels with the same level of depth changes are represented with a layer to enable efficient S-VSD calculation at the layer level. Meanwhile, a nonlinear mapping function is learnt to accurately represent the relationship between the VSD and S-VSDs, automatically providing weights for S-VSDs during the VSD estimation. To learn such function, a dataset of VSD and its associated S-VSDs are built. Experimental results show that the VSD can be accurately estimated with the weights learnt by the nonlinear mapping function once its associated S-VSDs are available. The proposed method outperforms the relevant state-of-the-art methods in both accuracy and efficiency. The dataset and source code of the proposed method will be available at https://github.com/jianjin008/.

\end{abstract}

%
\IEEEpeerreviewmaketitle

\section{Introduction}
\label{Intro}
%
%
%
%
%
\begin{figure}[htbp]
	\begin{center}
		\noindent
		\includegraphics[width = 3.4in]{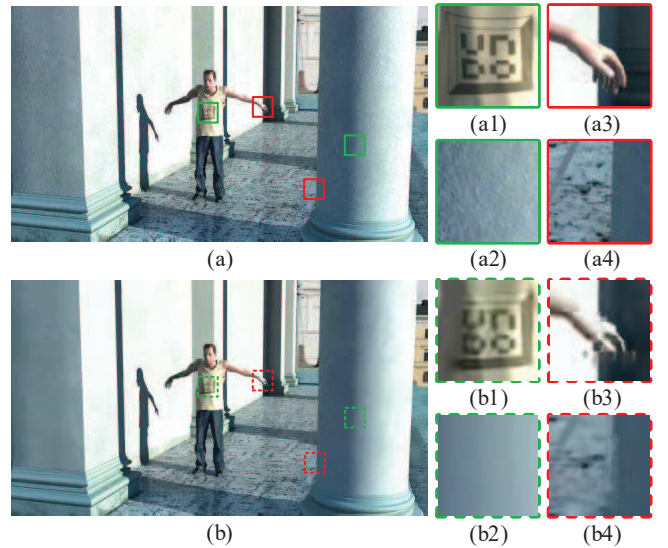}
		\caption{Illustration of lossy compression caused texture degradation with different levels of position shifting distortion. (a) and (b) are synthesized by original reference views and decoded reference views, respectively. The green patches located on the main bodies of person and cylinder hardly have depth boundaries, where only small depth changes caused by compression exist and lead to small position shifting in (b1) and (b2). The red patches have lots of depth boundaries around the fingers and cylinder. Compression causes large depth changes around depth boundaries, leading to large position shifting, e.g., finger misalignment and cylinder boundary erosion in (b3) and (b4). Besides, texture changes in the decoded texture reference views are propagated to the virtual view under the direction of changed depth, causing texture degradation from (b1) to (b4).}\label{Intro}
	\end{center}
	\vspace{-0.5cm}
\end{figure}
\subsection{Motivation}
In recent years, 3-D video technology has been popular due to its fresh viewing experiences, e.g., special immersion, high interactivity, and large degree of freedom. In the 3-D video system, the multiview view plus depth (MVD) \cite{merkle2007multi} representation is the main data format. The MVD records the color and depth information of the same physical scene from different views. With the MVD format data, arbitrary virtual views can be synthesized via a depth-image-based rendering (DIBR) technique \cite{fehn2004depth,jin2016region}. Commonly, the performance of 3-D video system is mainly measured by the distortion/quality \cite{tian2017niqsv+,tian2018benchmark,ling2020re} of the synthesized virtual view. Hence, the view synthesis distortion (VSD) estimation is crucial, especially for the 3-D video applications. For instance, the estimated VSD \cite{zhang2013regional} is generally used for rate-distortion optimization \cite{yuan2011model}, rate allocation \cite{shao2011asymmetric}, the design of the error resilience techniques \cite{gao2014rate}, etc.

The main reason for the VSD is the changes/errors in the reference texture and depth videos due to lossy compression or transmission errors. During the view synthesis process, texture changes may cause VSD in the luminance/chrominance level, namely texture degradation. However, depth changes may cause complex geometric VSD. Moreover, different levels of depth changes lead to different levels of geometric VSD. In 1-D parallel model, the geometric VSD commonly refers to position shifting \cite{zheng2017fine}. Besides, texture degradation propagating from decoded reference views to their corresponding virtual views is also directed by the changes of depth. After integrating texture degradation with the position shifting, the texture degradation with different levels of position shifting can be regarded as different kinds of sub-VSD (S-VSD) and forms the final VSD.

As shown in Fig. \ref{Intro}, (a) and (b) are synthesized by original reference views (uncompressed in texture and depth reference views) and decoded reference views (compressed by H.264 with QP pair (45, 48) in texture and depth reference views), respectively. Some magnification patches of local distortion are exhibited in (a1)-(a4) and (b1)-(b4). The VSD in the green/red patches mainly belongs to the texture degradation with small/large position shifting, as shown in (b1-b2)/(b3-b4), where the shifting is mainly due to the depth changes caused by lossy compression. As the green patches locate on the bodies of person and cylinder, their original depth is smooth. Even after compressed, the level of depth changes in green patches is low, which only causes non-obvious position shifting. After integrating with the texture degradation, only texture degradation is obviously observed in the green patches. However, the VSD in the red patches mainly belongs to the texture degradation with large position shifting, such as the misaligned fingers in (b3) and the erosion around the boundary of cylinder in (b4). Since obvious depth boundaries exist around the hand and cylinder, lossy compression makes these boundaries smooth and brings large level of depth changes. This leads to the significant position shifting effects in the red patches. After texture degradation propagation, obvious texture degradation together with position shifting can be observed in the green patches. All these distortion, including that in (b1)-(b4), forms the final VSD in (b). 

Inspired from above, after taking the texture distortion into account, different levels of depth changes can be used to represent different kinds of S-VSD, which can be further used to predict the VSD. On the one hand, it can benefit the optimization of 3-D video coding \cite{zhao2015scalable} by figuring out the exact contribution of each kind of S-VSD to the VSD. On the other hand, it can also help us design an optimal depth codec \cite{jin2015accurate} by increasing or decreasing different levels of depth changes to bring in the smallest VSD. To our best knowledge, existed methods, such as the methods to be reviewed in subsection I-C-\emph{2)}, cannot represent the relationship between the S-VSD and VSD accurately, which is the key challenge in this work.

\subsection{Our contributions}
In this paper, we propose an auto-weighted layer representation based view synthesis distortion estimation model. 
This is the first work utilizing learning-based approach to mine the accurate relationships among the degeneration of texture, changes of depth, and the VSD, especially for the relationship between the VSD and its associated S-VSDs. This provides us a methodology to predict the VSD by utilizing its associated S-VSDs. It can be used for optimizing the design of 3-D video coding, especially for the depth coding. The main contributions are summarized as follows.

\begin{itemize}
    \item This is the first work to relate different levels of depth changes together with their texture degeneration to the view synthesis distortion (VSD), which is crucial for various 3-D video applications, such as aforementioned 3-D video coding, depth coding, etc.  
    
    \item Sub view synthesis distortion (S-VSD) is first defined according to the level of depth changes and its associated texture degeneration in this paper. Besides, an elaborate derivation is given to demonstrate that the VSD can be decomposed into different kinds of S-VSD approximately. 
    
    \item To accurately represent the relationship between VSD and its associated S-VSDs, a nonlinear mapping function between the VSD and S-VSDs is learnt based on our built dataset, which is the first dataset for mining the relationship between the VSD and S-VSDs.
    
    \item To calculate the S-VSD efficiently, a layer-based representation method is proposed and further optimized, where all the pixels with the same level of the depth changes (i.e., the S-VSD) will be represented with a layer. It enables the S-VSD calculation perform at the layer level.
\end{itemize}

Compared with the existed VSD estimation methods, the well-learnt nonlinear mapping function is able to accurately represent the relationship between the VSD and S-VSDs. Meanwhile, the proposed layer-based representation enables the VSD estimation performed at the layer level without spending additional calculation on partly performing the view synthesis process at the pixel level to make the proposed method more efficient.

\subsection{Related work}
\subsubsection{View synthesis} In this paper, View synthesis mainly refer to the DIBR based view synthesis, which commonly contains two steps, namely warping and blending. 

During the warping step, forward warping, warping competition, and rounding operation are performed accordingly. The goal of warping step is to warp the pixels in the reference views to the warped views. Assume that a pixel in the original reference view with location $(i,j_k)$ is warped to a new location $(i,j)$ in the warped view. The subscript $k$ is used to index the left ($k=0$) or right ($k=1$) view. This process can be formulated as
\begin{equation}
\begin{split}
\label{dis}
&\phi_k=j-j_{k}=\left[ \frac{f B_{k} D(i, j_{k})}{255}\left(\frac{1}{Z_{\text {near}}}-\frac{1}{Z_{\text {far}}}\right)+\frac{f B_k}{Z_{\text {far}}}\right]\\
& =\Phi(D(i,j_k)),
\end{split}
\end{equation}     
where $\phi_k$ denotes the disparity of pixel $(i,j_k)$ with depth value $D(i,j_k)$. $\left[ \cdot \right]$ is the rounding operation. $B_k$ denotes the baseline between cameras and $f$ is the focal length of cameras.  $[Z_{\text {near}}, Z_{\text {far}}]$ is the depth range of the physical scene. In Eq. \eqref{dis}, disparity $\phi_k$ can be regarded as the function of depth value $D(i,j_k)$, which is denoted as $\Phi(D(i,j_k))$ for simplification.   


After warping step, there will be lots of dis-occlusions in the warped views, since the occlusion parts in the reference views become visible. To fill the dis-occlusions, the blending step is carried out by merging the two warped views into a virtual one. Besides, three blending strategies need to be followed according to three different cases during blending step: i) if current pixel in the virtual view is visible in both two warped views, a weighted average of these two values is used; ii) if it is only visible in one of the warped views, this value will be directly used; iii) otherwise, an inpainting value will be used. 

\subsubsection{View synthesis distortion estimation} Commonly, there are two typical VSD estimation methods to estimate lossy compression caused view synthesis distortion and transmission error (package loss) caused view synthesis distortion, respectively. Kim $et$ $al.$ \cite{kim2009depth} first developed a camera and video parameters based quality metric to quantify the effect of lossy coding of depth maps on synthesized view quality. After that, Yuan $et$ $al.$ \cite{yuan2011model} proposed a concise distortion model by analyzing the impacts of the compression distortion of texture images and depth maps on the quality of the virtual views. Meanwhile, Zhang $et$ $al.$ \cite{zhang2013regional} proposed a view synthesis distortion model by taking the regions characteristics into account for depth video coding. 
Based on this, Fang $et$ $al.$ \cite{fang2013analytical} relate errors in the depth images to the synthesis quality by taking texture image characteristics and the warping step of view synthesis into account. However, the warping step is used for relating the distortion of depth to the synthesized view at the frame level, which limits its accuracy on the VSD estimation. To make more accurate prediction of the VSD, Yuan $et$ $al.$ \cite{yuan2016virtual} utilize the warping step of view synthesis to simulate the error prorogation from distorted depth to the virtual synthesized view at the pixel level, which directly measures the quality of the virtual view by partly carrying out view synthesis. However, the blend and inpainting steps in view synthesis are still not considered due to their complicated operations. Jin $et$ $al.$ \cite{jin2019pixel} proposed a pixel-level VSD estimation, where the warping and blending steps are partly taken into account to build a more accurate relations between the distorted depth together with the texture and VSD, achieving the state-of-the-art result. However, compared with the pixel-level VSD estimation methods in \cite{yuan2016virtual} and \cite{jin2019pixel}, the frame-level one in \cite{fang2013analytical} is more efficient, when pixel-level parallel processing is not considered. Meanwhile, Pan $et$ $al.$ \cite{gao2019occlusion} developed a depth distortion range, in which depth changes brought no geometrical distortions. 

To model the transmission error caused distortion, Zhou $et$ $al.$ \cite{zhou2011channel} first derived a Channel distortion model for multi-view video transmission over lossy packet-switched networks, which can estimate the channel caused distortion at the frame level. Then, a quadratic model is proposed by Cheung $et$ $al.$ \cite{cheung2011transform}, which first relates the disparity errors caused by packet loss in the depth maps to the distortion contribution in the synthesized view. After that, Gao $et$ $al.$ \cite{gao2014rate} developed an end-to-end 3-D video transmission oriented VSD estimation model for 3-D video coding to improve error resilience. To accurately model the error propagation process during view synthesis, Zhang $et$ $al.$ \cite{zhang2014view} proposed a depth-value-based graphical (DVGM) model. By taking the transmission error into account, it can accurately estimate the transmission caused view synthesis distortion. To further speed up DVGM, Jin $et$ $al.$ \cite{jin2018depth} proposed a depth-bin based graphical model for VSD estimation, which is more efficient without sacrificing accuracy.

As reviewed above, all these methods are trying to predict the VSD by modeling frame-level depth distortion or pixel-level depth distortion without considering the exact contribution of different levels of depth changes for the VSD. Besides, to build the relations between the distorted reference views to the virtual synthesized view, all these VSD methods partly integrate the view synthesis process into their approaches, e.g. the warping step. This will lead to two drawbacks: 1) It cannot achieve accurate prediction of the VSD by partly using view synthesis (e.g., warping step) for building the relationship between the distorted texture together with depth and the VSD, since the blending step of the view synthesis also affects the view synthesis results. While the nonlinear operations of blending step (warping competition, inpainting operation, etc.) make it hardly formulated in such VSD estimation methods. 2) Even through part of view synthesis is performed in such methods, the calculation is based on pixel-level, namely each pixel will perform the view synthesis process partly. This reduces their efficiency in some degree. To overcome these drawbacks, we firstly learn the nonlinear mapping function based on our dataset to exploit an exact relationship between different levels of depth changes together with their associated distorted texture (the S-VSDs) and the VSD. Then, we propose an efficient layer-based representation method, which enables the VSD estimation performed at the layer level.      

The outline of the rest of our paper is as follows. First, the proposed model is presented in Section \ref{Ours}. Then, the layer-based representation method is developed in Section \ref{layer-based representation}. Section \ref{Exp} presents experimental results and Section \ref{Con} concludes this paper.

\section{The proposed Model}
\label{Ours}
In this section, we first define the total view synthesis distortion (VSD) and the sub view synthesis distortion (S-VSD) abstractly. To better understand the S-VSD, an analysis on the S-VSD is given in detail. After that, a set of theoretical derivations is given according to view synthesis process, from which we demonstrate that the VSD can be decomposed into the S-VSDs by their associated weights approximately. Finally, a nonlinear mapping function represented with neural networks is used to learn the weights between the VSD and its associated S-VSDs.  


\subsection{Definition of the VSD and S-VSD}
\label{S-II-A}     
In this paper, the view synthesis distortion of the virtual view (i.e., VSD) is formulated with the Mean Square Error (MSE) over the entire frame of synthesized view as used in \cite{fang2013analytical}, where 
 \begin{equation}
\label{MSE}
\text{MSE}=\frac{1}{W \cdot H} \cdot \sum_{i=0}^{W-1} \sum_{j=0}^{H-1} \left(T(i, j)-\tilde{T}(i, j)\right)^{2},
\end{equation}
where $W$ and $H$ denote the width and height of the virtual view. $T(i, j)$ and $\tilde{T}(i, j)$ are the color values of two pixels in two different virtual views, which are synthesized with the original reference views and decoded views. Besides, these two pixels have the same location $(i,j)$ in their own virtual views.  

As mentioned in Eq. \eqref{dis}, depth changes may lead to disparity changes. Assume that a pixel in the decoded reference view with location $(i,\tilde{j_k})$ is warped to $(i,j)$ in the warped view. Then, we have
\begin{equation}
\label{dispR}
\tilde{\phi}_k=j-\tilde{j}_{k}=\Phi(\tilde{D}(i,\tilde{j}_k)).
\end{equation}	
\begin{figure*}[htbp]
	\begin{center}
		\noindent
		\includegraphics[width = 7.1in]{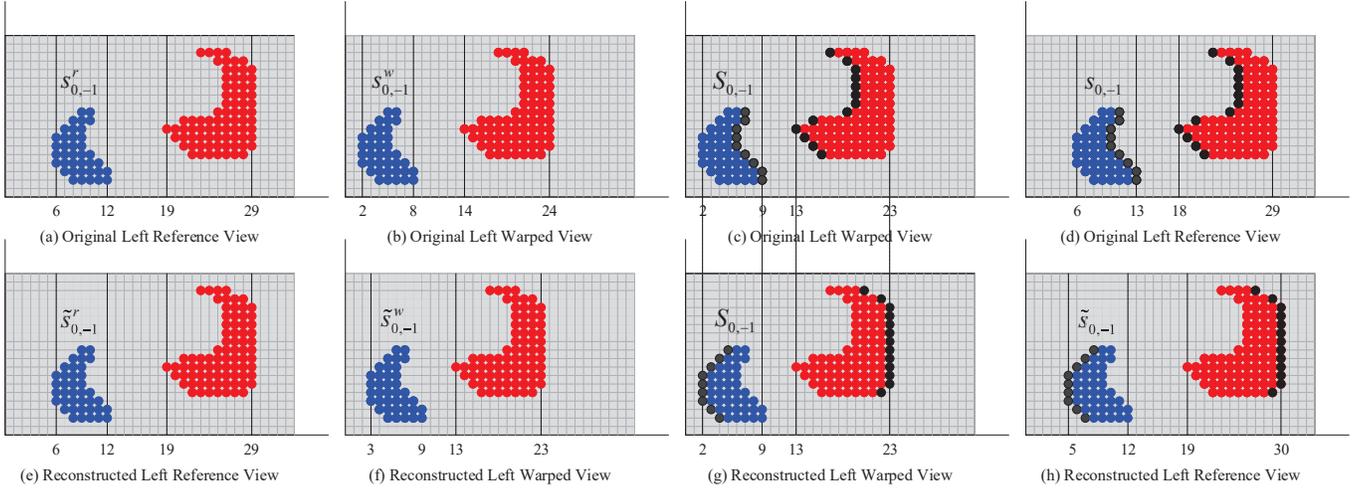}
		\caption{Illustration of the analysis on S-VSDs during view synthesis.}\label{Fig. 2}
	\end{center}
	\vspace{-0.5cm}
\end{figure*}
Besides, depth changes with different levels may cause disparities changes with different levels, namely warped position shifting with different levels. Here, the depth changes from $D(i,j_k)$ to $\tilde{D}(i,\tilde{j}_k)$ caused warped position shifting distortion is measured with $\Delta \phi_k$, and
\begin{equation}
\begin{split}
\label{shift}
& \Delta \phi_k = j_k - \tilde{j}_k = \tilde{\phi}_k - \phi_k = \Phi(\tilde{D}(i,\tilde{j}_k)) - \Phi(D(i,j_k)).
\end{split}
\end{equation}
Therefore, the warped position shifting distortion of a pixel can be functioned by the levels of depth changes from $D(i,j_k)$ to $\tilde{D}(i,\tilde{j}_k)$. After integrating with texture information distortion, the sub view synthesis distortion (S-VSD) in this paper is defined as
\begin{equation}
\small
\begin{split}
\label{S-VSD}
&L_{k, \Delta \phi_k} =\frac{1}{C_{k, \Delta \phi_k}} \sum_{(i, j_k) \in s_{k, \Delta \phi_k} \atop (i, \tilde{j}_k) \in \tilde{s}_{k, \Delta \phi_k}} \left ( T_{k}^{r}(i, j_k) - \tilde{T}_{k}^{r}(i, \tilde{j}_k) \right)^{2}\\
&=\frac{1}{C_{k, \Delta \phi_k}} \sum_{(i, j) \in S_{k, \Delta \phi_k}} \left ( T_{k}^{r}(i, j-\phi_{k}) - \tilde{T}_{k}^{r}(i, j-\tilde{\phi}_{k}) \right)^{2}\\
&=\frac{1}{C_{k, \Delta \phi_k}} \sum_{(i, j) \in S_{k, \Delta \phi_k}} \left ( T_{k}^{r}(i, j-\phi_{k}) - \tilde{T}_{k}^{r}(i, j-\phi_{k}-\Delta \phi_k) \right)^{2},\\ 
\end{split}
\end{equation}
where $L_{k, \Delta \phi_k}$ denotes a certain S-VSD of all the pixels $(i,j)$ that are with the same level of depth changes (measured with $\Delta \phi_k$) in the left ($k=0$) or right ($k=1$) warped views, and their locations are collected with set $S_{k,\Delta \phi_k}$. $C_{k, \Delta \phi_k}$ denotes the cardinality of $S_{k,\Delta \phi_k}$. Their associated pixels in the original and decoded reference views are $(i,j_k)$ and $(i,\tilde{j}_k)$, and their locations are collected with set $s_{k, \Delta \phi_k}$ and $\tilde{s}_{k, \Delta \phi_k}$, respectively. $T_{k}^{r}(i, j_k)$ and $\tilde{T}_{k}^{r}(i, \tilde{j}_k)$ are the texture value of pixel $(i,j_k)$ and $(i,\tilde{j}_k)$. It should be noticed that the whole frame collected with set $S_k$ can be divided into several $S_{k,\Delta \phi_k}$ according to $\Delta \phi_k$, and we have $S_k = S_{k,-3\sigma}\cup S_{k,-3\sigma+1} \cup\cdots\cup S_{k,3\sigma}$ and $C_k = W \cdot H$. Besides, the intersection of any two sets of $S_{k,\Delta \phi_k}$ is empty, i.e., $ S_{k,p} \cap S_{k,q}=\emptyset, \text{where}\ p \neq q\  \text{and}\ p,q \in [-3\sigma,3\sigma]$. 

As the lossy compression (source coding) caused depth distortion can be approximately regarded as a zero-mean white noise, the three-sigma rule is used to confirm the available number of $\Delta \phi_k$, namely $\Delta \phi_k \in [-3\sigma,3\sigma]$. 

\subsection{An detailed analysis of the S-VSD}
\label{II-B}
To better understand S-VSD, an analysis of the S-VSD is given based on an example in this subsection. Assume that there are three different levels of depth changes, caused by the lossy compression in the left and right depth reference views. Therefore, six S-VSDs need to be calculated for each of the VSD. Here, only three S-VSDs in the left view are involved for simplify, namely $L_{0,-1}$, $L_{0,0}$, and $L_{0,1}$. The other three S-VSDs ($L_{1,-1}$, $L_{1,0}$, and $L_{1,1}$) in the right view are similar. 

As shown in Fig. \ref{Fig. 2}, pixels with the same warped position shifting distortion are masked with the same color points in original and decoded reference views. Pixels with $\Delta \phi_k=-1$ and $\Delta \phi_k=1$ are highlighted with blue and red points, respectively. The rest are pixels with $\Delta \phi_k=0$, which means no warped position shifting distortion exists in these pixels. Set $s^{r}_{0,-1}$ is used to collect locations of the blue points in the original reference view, which has a horizontal interval $[6,12]$. Then, these pixels are warped to the interval $[2,8]$ in the original warped view, their locations are collected with set $s^{w}_{0,-1}$. As lossy compression makes the depth values of these pixels changed, the pixels in set $\tilde{s}^{r}_{0,-1}$ in the decoded reference view are mistakenly warped to a new interval $[3,9]$ in the decoded warped view. Their locations are collected with set $\tilde{s}^{w}_{0,-1}$. Assume all the pixels of these two sets are eventually exhibited in the virtual views. By fusing these two sets with a union operation, we can obtain set $S_{0,-1}$, namely $S_{0,-1} = s^{w}_{0,-1} \cup \tilde{s}^{w}_{0,-1}$. The complementary parts are highlighted with the dark points. Finally, we can easily locate their associated sets $s_{0,-1}$ and $\tilde{s}_{0,-1}$ in the original and decoded reference views by backward warping \cite{li2018hole} pixels in the $S_{0,-1}$. Therefore, the S-VSD $L_{0,-1}$ can be regarded as the Mean Square Error over the pixels of these two sets physically. Likewise, the definitions for other S-VSDs in the left view are similar..

It should be noticed that the purpose of union operation is to ensure the distorted pixels could be fully counted during the S-VSD calculation. Besides, depending on the fact in the backward warping that the missing depth value of a certain pixel is assigned with its adjacent pixel's depth value, we assign the depth values of the dark points with their adjacent pixels in the sets during inverse warping process.   

\subsection{A theoretical derivation: from VSD to S-VSDs}   

To relate the pixel in the virtual view with its associated pixels in the original and decoded reference views, we also take the inverse process into account as in \cite{jin2019pixel}. First, we relate the pixel in the virtual view to its associated pixels in the warped views by taking the blending strategies into account. $T(i, j)$ in Eq. \eqref{MSE} can be rewritten as 
\begin{equation}
\label{region}
T(i, j)=\left\{\begin{array}{c}{u_0 \cdot T_{0}^{w}(i,j)+u_1 \cdot T_{1}^{w}(i,j)} \\ {T_{0}^{w}(i,j)} \\ {T_{1}^{w}(i,j)} \\ {\text {Inpainting}}\end{array}\right.,
\end{equation}
where $T_{0}^{w}(i,j)$ and $T_{1}^{w}(i,j)$ denote the color values of pixels in the original left and right warped views. $u_{0}$ and $u_{1}$ ($u_0\in(0,1)$, $u_1\in(0,1)$) are two weights for the left and right warped views, which are determined by the locations of cameras array. The first item represents that the pixel $(i,j)$ in the virtual view is visible in both warped views case. The second and third items represent that only one of the warped views can be visible. The last item is to formulate the inpainting case. 

Similarly, $\tilde{T}(i, j)$ can be rewritten as
\begin{equation}
\label{region2}
\tilde{T}(i, j)=\left\{\begin{array}{c}{u_0 \cdot \tilde{T}_{0}^{w}(i,j)+u_1 \cdot \tilde{T}_{1}^{w}(i,j)} \\ {\tilde{T}_{0}^{w}(i,j)} \\ {\tilde{T}_{1}^{w}(i,j)} \\ {\text {Inpainting}}\end{array}\right..
\end{equation}
where $\tilde{T}_{0}^{w}(i,j)$ and $\tilde{T}_{1}^{w}(i,j)$ denote the color values of pixels in the decoded left and right warped views.

Then, we make an approximate computation. As the last case is infrequent, which occupies around 1$\%$ of the all these four cases as mentioned in \cite{jin2016region}. Therefore, $T(i,j)$ can be approximately rewritten as
\begin{equation}
\label{regionR}
T(i, j) \approx v_0 \cdot T_{0}^{w}(i,j)+v_1 \cdot T_{1}^{w}(i,j),
\end{equation}
where $v_0$ and $v_1$ ($v_0 \in [0,1]$, $v_1 \in [0,1]$) are two weights for the left and right warped views. Similarly, we have
\begin{equation}
\label{region2R}
\tilde{T}(i, j) \approx v_0 \cdot \tilde{T}_{0}^{w}(i,j)+v_1 \cdot \tilde{T}_{1}^{w}(i,j).
\end{equation}

\newcounter{TempEqCnt1} 
\setcounter{TempEqCnt1}{\value{equation}} 
\setcounter{equation}{15} 
\begin{figure*}[htbp]
\begin{equation}
	\begin{split}
	\label{MSEnew}
	\text{MSE} &\approx \frac{1}{W \cdot H} \cdot \sum_{i=0}^{W-1} \sum_{j=0}^{H-1} \left( v_0 \cdot (T_0^r(i,j-\phi_0) - \tilde{T}_0^r(i,j-\tilde{\phi}_0) ) + v_1 \cdot (T_1^r(i,j-\phi_1) - \tilde{T}_1^r(i,j-\tilde{\phi}_1) ) \right)^{2}\\
	&= \frac{1}{W \cdot H} \cdot \sum_{k=0}^{1} \sum_{i=0}^{W-1} \sum_{j=0}^{H-1}  \left( v_k \cdot (T_k^r(i,j-\phi_k) - \tilde{T}_k^r(i,j-\tilde{\phi}_k)) \right)^{2}\\
	&\approx \frac{v_k^2}{W \cdot H} \cdot \sum_{k=0}^{1}  \sum_{i=0}^{W-1} \sum_{j=0}^{H-1} \left( T_k^r(i,j-\Phi(D(i,j_k))) - \tilde{T}_k^r(i,j-\Phi(D(i,\tilde{j}_k))) \right)^{2}\\
	&= \frac{v_k^2}{W \cdot H} \cdot \sum_{k=0}^{1}  \sum_{i=0}^{W-1} \sum_{j=0}^{H-1}  \left( T_k^r(i,j-\phi_k) - \tilde{T}_k^r(i,j-\phi_k-\Delta \phi_k) \right)^{2},~\text{where}~\Delta \phi_k \in [-3\sigma,3\sigma].\\
	\end{split}
\end{equation}
\end{figure*}
\setcounter{equation}{\value{TempEqCnt1}}

\newcounter{TempEqCnt2} 
\setcounter{TempEqCnt2}{\value{equation}}
\setcounter{equation}{19} 
\begin{figure*}[htbp]
	\begin{equation}
	\begin{split}
	\label{MSEnew2}
	\text{MSE} &\approx \frac{1}{W \cdot H} \cdot \sum_{k=0}^{1} v_k^2 \left(C_{k,-3\sigma} \cdot L_{k,-3\sigma} + C_{k,-3\sigma+1} \cdot L_{k,-3\sigma+1} + \cdots + C_{k,3\sigma} \cdot L_{k,3\sigma}\right)\\
	&= \frac{v_0^2 \cdot C_{0,-3\sigma}}{W \cdot H} \cdot L_{0,-3\sigma}+\frac{v_0^2 \cdot C_{0,-3\sigma+1}}{W \cdot H} \cdot L_{0,-3\sigma+1}+...+\frac{v_0^2 \cdot C_{0,3\sigma}}{W \cdot H} \cdot L_{0,3\sigma}+\\
	&~~~~\frac{v_1^2 \cdot C_{1,-3\sigma}}{W \cdot H} \cdot L_{1,-3\sigma}+\frac{v_1^2 \cdot C_{1,-3\sigma+1}}{W \cdot H} \cdot L_{1,-3\sigma+1}+...+\frac{v_1^2 \cdot C_{1,3\sigma}}{W \cdot H} \cdot L_{1,3\sigma}.\\
	\end{split}
	\end{equation}
\end{figure*}
\setcounter{equation}{\value{TempEqCnt2}}

\newcounter{TempEqCnt3} 
\setcounter{TempEqCnt3}{\value{equation}}
\setcounter{equation}{20} 
\begin{figure*}[htbp]
	\begin{equation}
	\begin{split}
	\label{MSEnew3}
	\text{MSE} = \Psi(L_{0,-3\sigma}, L_{0,-3\sigma+1}, \cdots , L_{0,3\sigma}, L_{1,-3\sigma}, L_{1,-3\sigma+1}, \cdots , L_{1,3\sigma}).\\
	\end{split}
	\end{equation}
	\hrulefill
\end{figure*}
\setcounter{equation}{\value{TempEqCnt3}}

After that, we relate these pixels in the warped views to their associated pixels in the reference views by considering of the warping step during view synthesis. Likewise, we assume pixels with position $(i,j)$ in the original left and right warped views are warped from pixels with position $(i,j_k)$ in the original left ($k=0$) or right ($k=1$) reference views. We have
\begin{equation}
\label{r}
T_{k}^{w}(i,j) = T_{k}^{r}(i,j_k),
\end{equation}
where $T_{k}^{r}(i,j_k)$ denotes the color value of pixel $(i,j_k)$ in the original left ($k=0$) or right ($k=1$) reference views. Its associated disparity $\phi_k$ can be represented by Eq. \eqref{dis}.

Similarly, we assume pixels with position $(i,j)$ in the decoded warped views is warped from pixel with position $(i,\tilde{j}_k)$ in the decoded reference views due to its distorted depth values $\tilde{\phi}(i,\tilde{j}_k)$. We have 
\begin{equation}
\label{2r}
\tilde{T}_{k}^{w}(i,j) = \tilde{T}_{k}^{r}(i,\tilde{j}_k),
\end{equation}
where $\tilde{T}_{k}^{r}(i,\tilde{j}_k)$ denotes the color value of pixel $(i,j_k)$ in the decoded left ($k=0$) or right ($k=1$) reference view. Its associated disparities $\tilde{\phi}_k$ can be represented by Eq. \eqref{dispR}.

Therefore, $T_{k}^{w}(i,j)$ and $\tilde{T}_{k}^{w}(i,j)$ can be rewritten as
\begin{equation}
\label{R}
T_{k}^{w}(i,j) = T_{k}^{r}(i,j_k) = T_{k}^{r}(i,j-\phi_k)
\end{equation}
and
\begin{equation}
\label{2R}
\tilde{T}_{k}^{w}(i,j) = \tilde{T}_{k}^{r}(i,\tilde{j}_k) = \tilde{T}_{k}^{r}(i,j-\tilde{\phi}_k).
\end{equation}
By substituting Eq. \eqref{R} and Eq. \eqref{2R} into Eq. \eqref{regionR} and Eq. \eqref{region2R}, respectively, we obtain 
\begin{equation}
\label{T1}
T(i,j) \approx v_0 \cdot T_{0}^{r}(i,j-\phi_0) + v_1 \cdot T_{1}^{r}(i,j-\phi_1)
\end{equation}
and 
\begin{equation}
\label{T2}
\tilde{T}(i,j) \approx v_0 \cdot \tilde{T}_{0}^{r}(i,j-\tilde{\phi}_0) + v_1 \cdot \tilde{T}_{1}^{r}(i,j-\tilde{\phi}_1).
\end{equation}
Eq. \eqref{T1} and Eq. \eqref{T2} relate the pixel in the virtual view to its associated pixels in the reference views. Finally, Eq. \eqref{MSE} is rewritten as Eq. \eqref{MSEnew}. 

For the Eq. \eqref{S-VSD}, it can be rewritten as 

\setcounter{equation}{16}
\begin{equation}
\begin{split}
\label{S-VSD2}
&\sum_{(i, j) \in S_{k, \Delta \phi_k}} \left ( T_{k}^{r}(i, j-\phi_{k}) - \tilde{T}_{k}^{r}(i, j-\phi_{k}-\Delta \phi_k) \right)^{2}\\
&=C_{k, \Delta \phi_k} \cdot L_{k, \Delta \phi_k}.\\
\end{split}
\end{equation}

As aforementioned that the intersection of any two sets of $S_{k,\Delta \phi_k}$ is empty and $S_k = S_{k,-3\sigma}\cup S_{k,-3\sigma+1} \cup\cdots\cup S_{k,3\sigma}$, we have

\begin{equation}
\begin{split}
\label{S-VSD3}
&\sum_{(i, j) \in S_{k}} \left ( T_{k}^{r}(i, j-\phi_{k}) - \tilde{T}_{k}^{r}(i, j-\phi_{k}-\Delta \phi_k) \right)^{2}\\
&=C_{k, -3\sigma} \cdot L_{k, -3\sigma}+C_{k, -3\sigma+1} \cdot L_{k, -3\sigma+1}+ \cdots\\ 
&~~~+ C_{k, 3\sigma} \cdot L_{k, 3\sigma}.\\
\end{split}
\end{equation}

As $C_k = W \cdot H$, then we obtain 
\begin{equation}
\begin{split}
\label{S-VSD4}
&\sum_{(i, j) \in S_{k}} \left ( T_{k}^{r}(i, j-\phi_{k}) - \tilde{T}_{k}^{r}(i, j-\phi_{k}-\Delta \phi_k) \right)^{2}\\
&=\sum_{i=0}^{W-1} \sum_{j=0}^{H-1} \left( T_k^r(i,j-\phi_k) - \tilde{T}_k^r(i,j-\phi_k-\Delta \phi_k) \right)^{2}.\\
\end{split}
\end{equation}

By plugging Eq. \eqref{S-VSD3} and Eq. \eqref{S-VSD4} into Eq. \eqref{MSEnew}, we have Eq. \eqref{MSEnew2}. Based on the observation of Eq. \eqref{MSEnew2}, the VSD can be decomposed into S-VSDs ($L_{k,\Delta \phi_k}$) by their associated weights approximately. To obtain the exact weights between the VSD and its associated S-VSDs, 
a nonlinear mapping function represented with neural networks is used instead of the linear one to well learn the non-linear relation between the VSD and S-VSDs due to a nonlinear mapping represented with neural networks is employed instead of a linear one to well learn the non-linear relation between the VSD and S-VSDs. This is due to the non-linear operations existed in view synthesis, e.g., hole filling, inpainting, warping competition and so on.
Then, we have Eq. \eqref{MSEnew3}, where the VSD is regarded as a non-linear mapping function of S-VSDs via a functional relation $\Psi(\cdot)$. This also gives us a theoretical proof that the VSD can be predicted using its associated S-VSDs once the functional relation $\Psi(\cdot)$ is obtained.

%

%
%

\subsection{Learning a non-linear mapping function: $\Psi (\cdot)$}

To make the prediction from S-VSDs closely approximate to the actual VSD, we formulate an optimization program as follows 
\setcounter{equation}{21}
\begin{equation}
\label{Objective_v1}
\min_{\Psi} \sum_{n=1}^{N} f \left(  \Psi \left(L_{k, \Delta \phi_k, n} \right), \text{MSE}_{n} \right), 
\end{equation}
where $n$ is the index of sample, and $N$ training samples are involved. $L_{k, \Delta \phi_k, n}$ denotes a serious S-VSDs (i.e., $L_{0, -3\sigma, n}, \cdots, L_{0, 3\sigma, n}, L_{1, -3\sigma, n}, \cdots, L_{1, 3\sigma, n} $) of sample $n$, which can be obtained by our layer-based representation method, which is introduced in Section~\ref{layer-based representation}. Their associated actual VSD (i.e., MSE$_n$) can be directly obtained by calculating the mean square error between the $n$ pair of virtual views, which are synthesized by their corresponding original reference views and decoded reference views. $f(\cdot,\cdot)$ is used to measure the approximation between two terms. With such a learnt nonlinear mapping function $\Psi(\cdot)$, once another series of S-VSDs are given, their associated VSD can be accurately predicted correspondingly. 

In particular, to obtain such a nonlinear function $\Psi(\cdot)$, we employ XGBoost \cite{chen2016xgboost}, a scalable end-to-end tree boosting system. Specifically, by introducing $M$ tress to our model, the optimization program in Eq. \eqref{Objective_v1} can be rewritten as

\begin{equation}
\begin{split}
\label{Objective_xgboost}
\min_{\substack{\Psi_{m}\\m=1,\cdots,M}} &\sum_{n=1}^{N} g \left(  \sum_{m=1}^{M} (\Psi_{m} \left(L_{k, \Delta \phi_k, n} \right)), \text{MSE}_{n} \right)\\ 
& + \sum_{m=1}^{M} \Omega(\Psi_{m}),
\end{split}
\end{equation}
\begin{figure*}[htbp]
	\begin{center}
		\noindent
		\includegraphics[width = 7.1in]{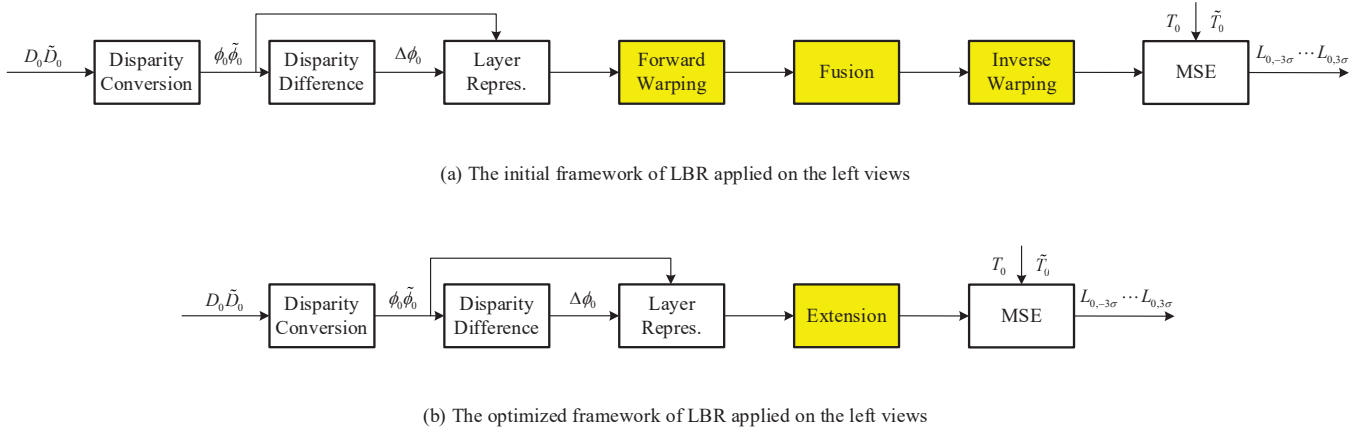}
		\caption{Illustration of the framework of our proposed layer-based representation method.}\label{Fig. 3}
	\end{center}
	\vspace{-0.5cm}
\end{figure*}
\begin{figure}[htbp]
	\begin{center}
		\noindent
		\includegraphics[width = 3.1in]{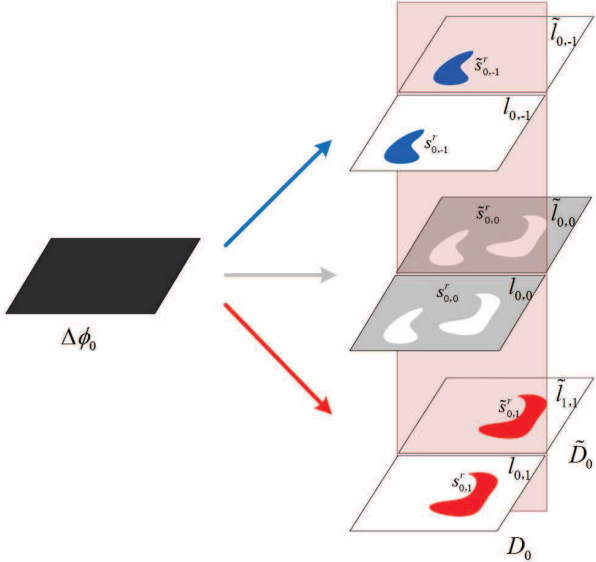}
		\caption{Illustration of layered representation. }\label{Fig. 4}
	\end{center}
	\vspace{-0.5cm}
\end{figure}
where $g(\cdot,\cdot)$ represents the training error of $n^{th}$ sample, and we adopt the mean square error. The second term measures the complexity of our tree, where $\Omega(\Psi_{m})$ is the complexity of $m^{\text{th}}$ tree.

Of note, minimizing the tree complexity facilitates the generalization ability of our algorithm, while minimizing the prediction error can guarantee the accuracy of our model. The parameters setting of XGBoost will be given in subsection IV-A.

%

%

\section{Layer-Based Representation}
\label{layer-based representation}
After analyzing the S-VSD, a layer-based representation method is first developed to generate the S-VSD in this section. Then, an optimization of the layer-based representation is made to reduce its complexity and speedup the S-VSDs generation. The details will be presented as follows.

\subsection{Methodology}
It is similar that the layer-based representation method is performed on the left and right views. For simplify, only its application on the left views is carefully elaborated in the following parts. The framework of the layer-based representation is shown in Fig. \ref{Fig. 3} (a). 

\subsubsection{Disparity Conversion}
The depth images are input at first, which contains the left original and decoded depth images $D_0$ and $\tilde{D}_0$. Then, $\phi(D_0)$ and $\phi(\tilde{D}_0)$ are obtained by plugging  $D_0$ and $\tilde{D}_0$ into Eq. \eqref{dis}, respectively. 

\subsubsection{Disparity Difference}
A pixel-wise subtraction operation is carried out to obtain the difference between $\phi(D_0)$ and $\phi(\tilde{D}_0)$. Then, we have the disparity difference image $\Delta \phi_0$.   

\subsubsection{Layered Representation}
Pixels with the same value in $\Delta \phi_0$ are masked with the same color and collected with a pair of layers $l_{0,\Delta \phi_0}$, $\tilde{l}_{0,\Delta \phi_0}$ in $D_0$ and $\tilde{D}_0$, respectively. We use different colors to mask pixels with different values, which are further represented with different layers. Then, sets $s^r_{0,-3\sigma}$, ... , $s^r_{0,3\sigma}$ and $\tilde{s}^r_{0,-3\sigma}$, ... , $\tilde{s}^r_{0,3\sigma}$ are easily obtained by visiting their corresponding layers $l_{0,\Delta \phi_0}$, $\tilde{l}_{0,\Delta \phi_0}$. An example with assumption $\sigma=1$ is shown in Fig. \ref{Fig. 4}, pixels with $\Delta \phi_0 = -1, 0, 1$ are masked with blue, gray, and red colors, which are represented with three layers in $D_0$ and $\tilde{D}_0$, respectively, i.e., $l_{0,-1}, l_{0,0}, l_{0,1}$, and $\tilde{l}_{0,-1}, \tilde{l}_{0,0},\tilde{l}_{0,1}$, which contains sets $s^r_{0,-1}, s^r_{0,0}, s^r_{0,1}$, and $\tilde{s}^r_{0,-1}, \tilde{s}^r_{0,0}, \tilde{s}^r_{0,1}$. It should be noticed that all the following operations are performed at the layer level.   

\subsubsection{Forward Warping}

The layered pixels in $D_0$ and $\tilde{D}_0$ are forward warped to the original and decoded warped views according to the disparity images $\phi(D_0)$ and $\phi(\tilde{D}_0)$. Sets $s^w_{0,-3\sigma}$, ... , $s^w_{0,3\sigma}$ and $\tilde{s}^w_{0,-3\sigma}$, ... , $\tilde{s}^w_{0,3\sigma}$ are obtained and represented with different pairs of layers.

\subsubsection{Fusion}
Merge each $s^w_{0,\Delta \phi_0}$ and $\tilde{s}^w_{0,\Delta \phi_0}$ (where $\Delta \phi_0 \in [-3\sigma,3\sigma]$) by a union operation and obtain $S_{0,-3\sigma}$, ... , $S_{0,3\sigma}$, namely $S_{0,\Delta \phi_0} = s^w_{0,\Delta \phi_0} \cup \tilde{s}^w_{0,\Delta \phi_0}$ (where $\Delta \phi_0 \in [-3\sigma,3\sigma]$). 

\subsubsection{Inverse Warping}
Inversely warp the $S_{0,\Delta \phi_0}$ back to the original and decoded left reference view, and generate their associated sets $s_{0,\Delta \phi_0}$ and $\tilde{s}_{0,\Delta \phi_0}$, which are presented with different pairs of layers. 

\subsubsection{MSE}
Pixels with locations $s_{0,\Delta \phi_0}$ in $T_0$ and $\tilde{s}_{0,\Delta \phi_0}$ in $\tilde{T}_0$ are used to calculate the $L_{0,\Delta \phi_0}$ via a layer level MSE calculation, which is similar with that used in Eq. \eqref{S-VSD}. 

\begin{table*}[htbp]
  \centering
  \caption{Details of Test Sequences}
    \begin{tabular}{l|c|c|c|c}
    \Xhline{1.2pt}
    Sequences & Resolutions & Views & Frames & QP Pairs (texture, depth) \\
    \Xhline{1.2pt}
    BookArrival \cite{HHI3d} & 1024*768 & 9~(8,10) & 1$^{th}$ to 25$^{th}$ & (15, 24), (20, 29), (25, 34), (30, 39), (35, 42),(40, 45),(45, 48) \\
    Kendo \cite{Nagoya3d} & 1024*768 & 2~(1,3) & 1$^{th}$ to 25$^{th}$ & (15, 24), (20, 29), (25, 34), (30, 39), (35, 42),(40, 45),(45, 48) \\
    Balloons \cite{Nagoya3d} & 1024*768 & 2~(1,3) & 1$^{th}$ to 25$^{th}$ & (15, 24), (20, 29), (25, 34), (30, 39), (35, 42),(40, 45),(45, 48) \\
    NewsPaper \cite{Newspaper} & 1024*768 & 3~(2,4) & 1$^{th}$ to 25$^{th}$ & (15, 24), (20, 29), (25, 34), (30, 39), (35, 42),(40, 45),(45, 48) \\
    PoznanStreet \cite{domanski2009poznan} & 1920*1080 & 4.5~(4,5) & 1$^{th}$ to 25$^{th}$ & (15, 24), (20, 29), (25, 34), (30, 39), (35, 42),(40, 45),(45, 48) \\
    PoznanHall2 \cite{domanski2009poznan} & 1920*1080 & 6.5~(6,7) & 1$^{th}$ to 25$^{th}$ & (15, 24), (20, 29), (25, 34), (30, 39), (35, 42),(40, 45),(45, 48) \\
    UndoDancer \cite{NokiaUndoDancer} & 1920*1080 & 3~(1,5) & 1$^{th}$ to 25$^{th}$ & (15, 24), (20, 29), (25, 34), (30, 39), (35, 42),(40, 45),(45, 48) \\
    GT-Fly \cite{NokiaGTFly} & 1920*1080 & 7~(5,9) & 1$^{th}$ to 25$^{th}$ & (15, 24), (20, 29), (25, 34), (30, 39), (35, 42),(40, 45),(45, 48) \\
    \Xhline{1.2pt}
    \end{tabular}%
  \label{tab:T2}%
  \vspace{-0.5cm}
\end{table*}%

\subsection{Optimization}
As analyzed in Fig. \ref{Fig. 2}, pixels with $s^r_{0,-1}$ in (a) are firstly warped to $s^w_{0,-1}$ in (b). Due to lossy compression of the depth image, pixels with the same location in (e) are mistakenly warped to $\tilde{s}^w_{0,-1}$ (f). From the (b) to (f), this process could be regarded as a left-shift-operation from $s^w_{0,-1}$ to $\tilde{s}^w_{0,-1}$, the shift-interval is one full pixel precision. To make sure that all these changed pixels are counted during our S-VSD calculations, dark points are complemented. Besides, these complemented pixels have the same depth value with their neighboring pixels. Thus, after inverse warping, the pixels with $s_{0,-1}$ in (d) can be regarded as a right extension of the pixels with $s^r_{0,-1}$ in (a), and the extended-interval is one full pixel precision. Similarly, the pixels with $\tilde{s}_{0,-1}$ in (h) can be treated as a left extension of the pixels with $\tilde{s}^r_{0,-1}$ in (e), and the extended-interval is one full pixel precision as well. According the observation above, the complicated forward warping, union operation, and inverse warping processes can be replaced by an extension process, which is a layer-level fusion operation. The optimized layer-based representation framework is shown in Fig. \ref{Fig. 3} (b).   

\subsubsection{For the left view, the extension process can be performed as follows}

If $\Delta \phi_0 < 0$, the $s_{0,\Delta \phi_0}$ in the $T_0$ can be generated by a right extension operation applied on the $s^r_{0,\Delta \phi_0}$, with $|\Delta \phi_0|$ full pixel precision extended-interval; The $\tilde{s}_{0,\Delta \phi_0}$ in the $\tilde{T}_0$ can be generated by a right extension operation applied on the $\tilde{s}^r_{0,\Delta \phi_0}$, with $|\Delta \phi_0|$ full pixel precision extended-interval; If $\Delta \phi_0 > 0$, the $s_{0,\Delta \phi_0}$ in the $T_0$ can be generated by a left extension operation applied on the $s^r_{0,\Delta \phi_0}$, with $|\Delta \phi_0|$ full pixel precision extended-interval; The $\tilde{s}_{0,\Delta \phi_0}$ in the $\tilde{T}_0$ can be generated by a left extension operation applied on the $\tilde{s}^r_{0,\Delta \phi_0}$, with $|\Delta \phi_0|$ full pixel precision extended-interval; Otherwise, the rest of the pixels in $T_0$ and $\tilde{T}_0$ are the $s_{0,\Delta \phi_0}$ and $\tilde{s}_{0,\Delta \phi_0}$. 

\subsubsection{For the right view, the extension process is opposite}

If $\Delta \phi_1 < 0$, the $s_{1,\Delta \phi_1}$ in the $T_1$ can be generated by a left extension operation applied on the $s^r_{1,\Delta \phi_1}$, with $|\Delta \phi_1|$ full pixel precision extended-interval; The $\tilde{s}_{1,\Delta \phi_1}$ in the $\tilde{T}_1$ can be generated by a left extension operation applied on the $\tilde{s}^r_{1,\Delta \phi_1}$, with $|\Delta \phi_1|$ full pixel precision extended-interval; If $\Delta \phi_1 > 0$, the $s_{1,\Delta \phi_1}$ in the $T_1$ can be generated by a right extension operation applied on the $s^r_{1,\Delta \phi_1}$, with $|\Delta \phi_1|$ full pixel precision extended-interval; The $\tilde{s}_{1,\Delta \phi_1}$ in the $\tilde{T}_1$ can be generated by a right extension operation applied on the $\tilde{s}^r_{1,\Delta \phi_1}$, with $|\Delta \phi_1|$ full pixel precision extended-interval; Otherwise, the rest of the pixels in $T_1$ and $\tilde{T}_1$ are the $s_{1,\Delta \phi_1}$ and $\tilde{s}_{1,\Delta \phi_1}$.

Therefore, instead performing the complicated forward warping, union operation, and inverse warping processes, a fusion-like operation is achieved at the layer level to generate the S-VSDs, which makes it more efficient.   

\section{Experimental Results}
\label{Exp}

In this paper, three state-of-the-art methods, namely Yuan\cite{yuan2016virtual}, Fang\cite{fang2013analytical}, and Jin\cite{jin2019pixel}, are chosen as the anchors in our comparisons. The training and testing data are firstly generated by calculating the VSD and its associated S-VSDs with the traditional MSE calculation and the proposed layer-based representation, respectively. After that, two main experiments are conducted, which contains: i) accuracy comparisons and ii) efficiency comparisons.

\subsection{Training and testing data generation}
Here, 8 test sequences from the Common Test Conditions (CTC) of the JCT-3V \cite{software} are used. There are left and right reference views for each sequence. For each of left or right reference view, there are texture and its associated depth videos. The first 25 frames of reference views are compressed with 7 QP pairs, which are further used to synthesize the virtual view. Therefore, There are 32 (8$\times$4) original videos (original left texture video, original left depth video, original right texture video, and original right depth video) and 224 (8$\times$4$\times$7) compressed videos. The details of these test sequences are exhibited in TABLE \ref{tab:T2}, which includes the sequences resolutions, view positions (where $x(y,z)$ denotes that the $x^{th}$ view is synthesized with the $y^{th}$ and $z^{th}$ views), the index of used frames, and recommended QP pairs for texture and depth videos.

To obtain the training and testing data, the VSD (ground truth) is firstly obtained by carrying out the MSE in Eq. \eqref{MSE} between the the original synthesized views and compressed synthesized views. They are synthesized with original reference views and compressed reference views, respectively. Then, 1400 (8$\times$7$\times$25) VSD results are obtained. After that, the proposed layer-based representation is conducted on the original and compressed reference views to obtain the S-VSDs. As aforementioned, three-sigma rule is used to confirm the available number of $\Delta \phi_k$ according to the depth distortion in different frames. We obtain 937 S-VSDs with $\Delta \phi_k$ = 1 and 463 S-VSDs with $\Delta \phi_k$ = 3. Then, the VSD and its associated S-VSDs are respectively divided into two parts, i.e., training and testing data. The ratio between training and testing data is 2:1. To be fair, the division is randomly performed three times and we get three groups of training and testing data. 

\begin{table*}[htbp]
	\centering
	\caption{The Comparison of Four Methods in Terms of The Prediction of MSE, PNSR, And Time Cost.}
	\begin{tabular}{lccccccc}
		\Xhline{1.2pt}
		& $\Delta \phi_k$\ (Frames)      & Method     & MSE   & $\Delta$MSE  & PSNR (dB) & $\Delta$PSNR (dB) & Time (s) \\
		\Xhline{1.2pt}
		\multicolumn{1}{c}{\multirow{15}[6]{*}{Test 1}} & \multirow{5}[2]{*}{$\Delta \phi_k=1$\ (313)} & GT    & 6.4743  & /     & 41.4288  & /     & / \\
		&       & Fang  & 7.6905  & 1.2162  & 40.3488  & 1.0800  & \ \textbf{\color{red}{0.6682 }} \\
		&       & Yuan  & 10.2094  & 3.7351  & 39.4684  & 1.9603  & 32.7745  \\
		&       & Jin & \textbf{\color{blue}{7.2745}}  & \textbf{\color{blue}{0.8002}}  & \textbf{\color{blue}{40.5898}}  & \textbf{\color{blue}{0.8390}}  & 3.3692  \\
		&       & Ours & \ \textbf{\color{red}{6.3341} } & \ \textbf{\color{red}{0.1402 }} & \ \textbf{\color{red}{41.4292 }} & \ \textbf{\color{red}{0.0004 }} & \textbf{\color{blue}{1.3349}}  \\
		\cline{2-8}          & \multirow{5}[2]{*}{$\Delta \phi_k=3$\ (155)} & GT    & 29.2056  & /     & 34.2465  & /     & / \\
		&       & Fang  & 30.4578  & 1.2522  & 33.9720  & 0.2745  & \ \textbf{\color{red}{0.5815 }} \\
		&       & Yuan  & 48.6340  & 19.4284  & 31.9915  & 2.2550  & 27.6567  \\
		&       & Jin & \textbf{\color{blue}{29.6714}}  & \textbf{\color{blue}{0.4657}}  & \textbf{\color{blue}{34.0862}}  & \textbf{\color{blue}{0.1603}}  & 3.0297  \\
		&       & Ours & \ \textbf{\color{red}{28.9729 }} & \ \textbf{\color{red}{0.2327 }} & \ \textbf{\color{red}{34.2446 }} & \ \textbf{\color{red}{0.0019 }} & \textbf{\color{blue}{1.7949}}  \\
		\cline{2-8}          & \multirow{5}[2]{*}{Average\ (468)} & GT    & 14.0029  & /     & 39.0500  & /     & / \\
		&       & Fang  & 15.2310  & 1.2281  & 38.2368  & 0.8132  & \ \textbf{\color{red}{0.6395 }} \\
		&       & Yuan  & 22.9355  & 8.9326  & 36.9921  & 2.0579  & 31.0795  \\
		&       & Jin & \textbf{\color{blue}{14.6923}}  & \textbf{\color{blue}{0.6894}}  & \textbf{\color{blue}{38.4358}}  & \textbf{\color{blue}{0.6142}}  & 3.2568  \\
		&       & Ours & \ \textbf{\color{red}{13.8320 }} & \ \textbf{\color{red}{0.1708 }} & \  \textbf{\color{red}{39.0497 }} & \ \textbf{\color{red}{0.0003 }} & \textbf{\color{blue}{1.4873}}  \\
		\Xhline{1.2pt}
		\multirow{15}[6]{*}{Test 2} & \multirow{5}[2]{*}{$\Delta \phi_k=1$\ (313)} & GT    & 6.6590  & /     & 41.3541  & /     & / \\
		&       & Fang  & 7.9168  & 1.2577  & 40.2716  & 1.0825  & \ \textbf{\color{red}{0.6687 }} \\
		&       & Yuan  & 10.5356  & 3.8766  & 39.3893  & 1.9649  & 32.9871  \\
		&       & Jin & \textbf{\color{blue}{7.4636}}  & \textbf{\color{blue}{0.8045}}  & \textbf{\color{blue}{40.4954}}  & \textbf{\color{blue}{0.8587}}  & 3.3976  \\
		&       & Ours & \ \textbf{\color{red}{6.5804 }} & \ \textbf{\color{red}{0.0786 }} & \ \textbf{\color{red}{41.3442 }} & \ \textbf{\color{red}{0.0100 }} & \textbf{\color{blue}{1.3389}}  \\
		\cline{2-8}          & \multirow{5}[2]{*}{$\Delta \phi_k=3$\ (155)} & GT    & 29.5211  & /     & 34.2191  & /     & / \\
		&       & Fang  & 30.7591  & 1.2379  & 33.9340  & 0.2852  & \ \textbf{\color{red}{0.5794 }} \\
		&       & Yuan  & 49.7187  & 20.1976  & 31.9193  & 2.2999  & 26.8840  \\
		&       & Jin & \textbf{\color{blue}{30.0273}}  & \textbf{\color{blue}{0.5062}}  & \textbf{\color{blue}{34.0644}}  & \textbf{\color{blue}{0.1548}}  & 3.0102  \\
		&       & Ours & \ \textbf{\color{red}{29.3131 }} & \ \textbf{\color{red}{0.2081 }} & \ \textbf{\color{red}{34.2078 }} & \ \textbf{\color{red}{0.0113 }} & \textbf{\color{blue}{1.7700}}  \\
		\cline{2-8}          & \multirow{5}[2]{*}{Average\ (468)} & GT    & 14.2309  & /     & 38.9911  & /     & / \\
		&       & Fang  & 15.4821  & 1.2512  & 38.1726  & 0.8184  & \  \textbf{\color{red}{0.6391}} \\
		&       & Yuan  & 23.5129  & 9.2820  & 36.9152  & 2.0758  & 30.9658  \\
		&       & Jin & \textbf{\color{blue}{14.9366}}  & \textbf{\color{blue}{0.7057}}  & \textbf{\color{blue}{38.3655}}  & \textbf{\color{blue}{0.6256}} & 3.2693  \\
		&       & Ours & \ \textbf{\color{red}{14.1094 }} & \ \textbf{\color{red}{0.1215 }} & \ \textbf{\color{red}{38.9806 }} & \ \textbf{\color{red}{0.0104 }} & \textbf{\color{blue}{1.4817}}   \\
		\Xhline{1.2pt}
		\multirow{15}[6]{*}{Test 3} & \multirow{5}[2]{*}{$\Delta \phi_k=1$\ (313)} & GT    & 7.0173  & /     & 41.1795  & /     & / \\
		&       & Fang  & 8.2793  & 1.2619  & 40.1180  & 1.0614  & \  \textbf{\color{red}{0.6528 }} \\
		&       & Yuan  & 11.1017  & 4.0844  & 39.2342  & 1.9453  & 31.7427  \\
		&       & Jin & \textbf{\color{blue}{7.8104}}  & \textbf{\color{blue}{0.7930}}  & \textbf{\color{blue}{40.3350}}  & \textbf{\color{blue}{0.8445}}  & 3.2600  \\
		&       & Ours & \ \textbf{\color{red}{6.9255 }} & \ \textbf{\color{red}{0.0918 }} & \ \textbf{\color{red}{41.1620 }} & \ \textbf{\color{red}{0.0175 }} & \textbf{\color{blue}{1.3047}}  \\
		\cline{2-8}          & \multirow{5}[2]{*}{$\Delta \phi_k=3$\ (155)} & GT    & 30.7228  & /     & 34.1629  & /     & / \\
		&       & Fang  & 31.9127  & 1.1899  & 33.8774  & 0.2856  & \ \textbf{\color{red}{0.5771 }} \\
		&       & Yuan  & 51.3203  & 20.5976  & 31.9013  & 2.2616  & 27.3320  \\
		&       & Jin & \textbf{\color{blue}{31.1067}}  & \textbf{\color{blue}{0.3839}}  & \textbf{\color{blue}{34.0060}}  & \textbf{\color{blue}{0.1570}}  & 2.9837  \\
		&       & Ours & \ \textbf{\color{red}{30.4047 }} & \ \textbf{\color{red}{0.3181 }} & \ \textbf{\color{red}{34.1480 }} & \ \textbf{\color{red}{0.0149 }} & \textbf{\color{blue}{1.7749}} \\
		\cline{2-8}          & \multirow{5}[2]{*}{Average\ (468)} & GT    & 14.8685  & /     & 38.8556  & /     & / \\
		&       & Fang  & 16.1066  & 1.2381  & 38.0511  & 0.8045  & \ \textbf{\color{red}{0.6277 }} \\
		&       & Yuan  & 24.4220  & 9.5535  & 36.8056  & 2.0500  & 30.2819  \\
		&       & Jin & \textbf{\color{blue}{15.5260}}  & \textbf{\color{blue}{0.6575}}  & \textbf{\color{blue}{38.2388}}  & \textbf{\color{blue}{0.6168}}  & 3.1685  \\
		&       & Ours & \ \textbf{\color{red}{14.7017 }} & \ \textbf{\color{red}{0.1668 }} & \ \textbf{\color{red}{38.8390 }} & \ \textbf{\color{red}{0.0166 }} & \textbf{\color{blue}{1.4604}}  \\
		\Xhline{1.2pt}
	\end{tabular}%
	\label{tab:All}%
	\vspace{-0.5cm}
\end{table*}%
%
\begin{table}[htbp]
	\centering
	\caption{Hyper Parameters Setting of XGBoost}
	\begin{tabular}{lc}
		\Xhline{1.2pt}
		Entries & Settings \\
		\Xhline{1.2pt}
		\textit{booster} & gbtree \\
		\textit{objective} & reg:gamma \\
		\textit{gamma} & 0.1 \\
		\textit{max\_depth} & 16 \\
		\textit{lambda} & 3 \\
		\textit{subsample} & 0.7 \\
		\textit{colsample\_bytree} & 0.7 \\
		\textit{min\_child\_weight} & 3 \\
		\textit{silent} & 1 \\
		\textit{eta} & 0.1 \\
		\textit{seed} & 1000 \\
		\textit{nthread} & 4 \\
		\Xhline{1.2pt}
	\end{tabular}%
	\label{tab:para}%
	\vspace{-0.5cm}
\end{table}%
\begin{figure*}[htbp]
	\begin{center}
		\noindent
		\includegraphics[width = 7.1in]{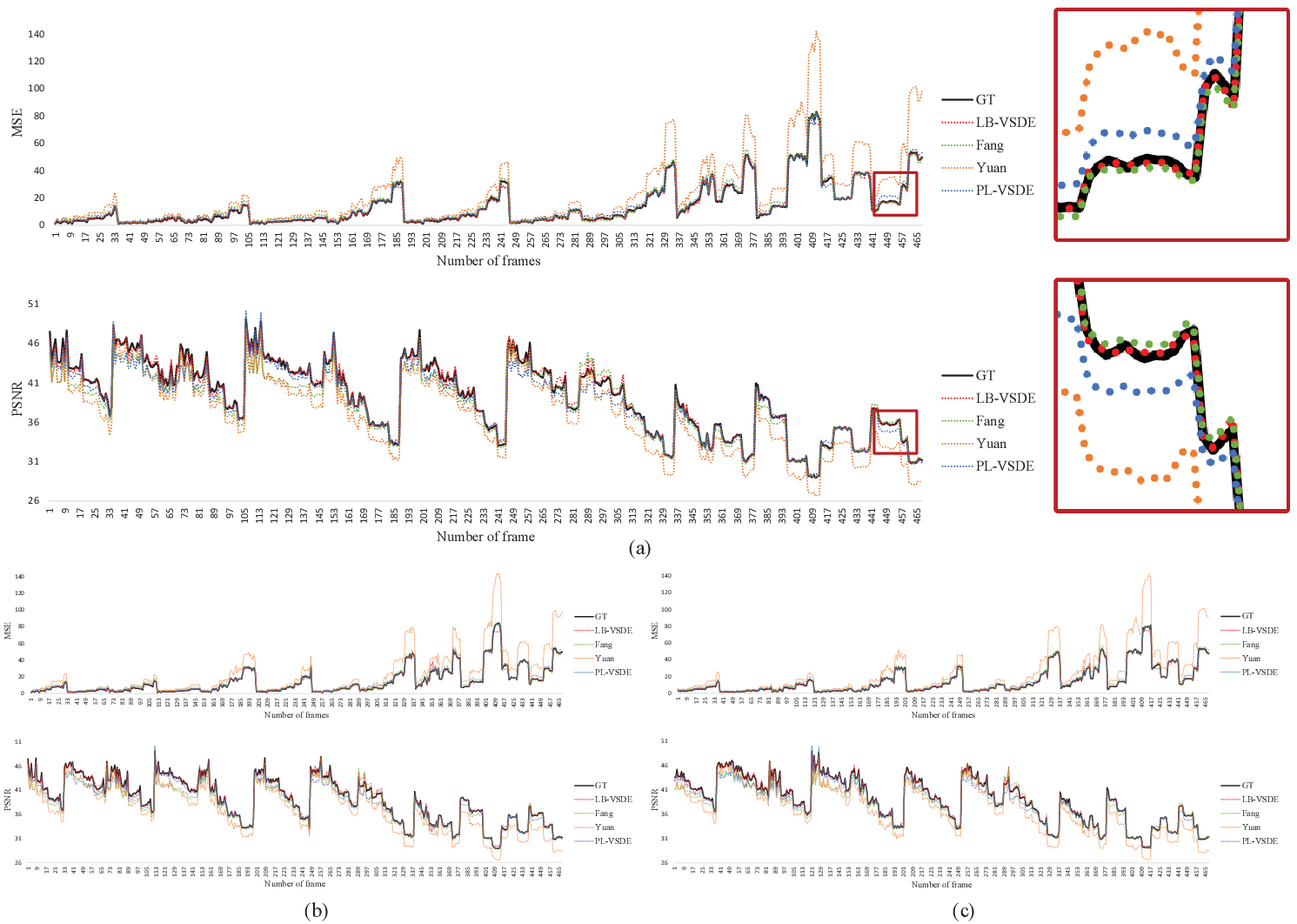}
		\caption{The comparison between the ground truth MSE \& PSNR and four predicted MSE \& PSNR provided by four different methods. (a), (b), and (c) are the results of three groups of training and testing data. The red patches are the magnification of local parts of the curve.}\label{Fig. 5}
	\end{center}
	\vspace{-0.5cm}
\end{figure*}

For each group of such data, we have 932 training data (624 training data with $\Delta \phi_k$ = 1 and 308 training data with $\Delta \phi_k$ = 3) and 468 testing data (313 testing data with $\Delta \phi_k$ = 1 and 155 testing data with $\Delta \phi_k$ = 3). The detailed settings and hyper parameters of XGBoost are shown in TABLE \ref{tab:para}. The training data are fed in the XGBoost system to train the nonlinear function $\Psi(\cdot)$. With the well learnt nonlinear function $\Psi(\cdot)$ and the testing data S-VSDs, VSD can be accurately predicted. In our experiments, three testing results are generated by training and testing on three groups of data.

\subsection{Accuracy comparison: MSE and PSNR}
As shown in TABLE \ref{tab:All}, the best and the second best results are highlighted with red and blue colors. Compared with three anchors, the proposed method achieves the best predicted results in both MSE and PSNR, i.e., achieving the smallest $\Delta$MSE and $\Delta$PSNR, where $\Delta$ is the absolute value of the difference between ground truth and predicted result. 

Besides, we also compare the ground truth MSE and PSNR with four predicted MSEs and PSNRs provided by four different methods in all testing frames. As shown in Fig. \ref{Fig. 5}, the experiments are conducted on three groups of training and testing data. Their results are shown in Fig. \ref{Fig. 5} (a), (b), and (c), respectively. The proposed method achieves the closest results with the ground truth in both MSE and PSNR.  

All these experimental results demonstrate that the well-learnt nonlinear mapping function can accurately represent the relationship between the VSD and its associated S-VSDs, which plays a critical role during view synthesis distortion estimation/prediction. With such well-learnt nonlinear mapping function, once the S-VSDs are given, their associated VSD can be accurately predicted in this work. On the one hand, it can facilitate the optimization of 3-D video coding by figuring out the exact contribution of each kind of S-VSD to the VSD. On the other hand, as the S-VSDs are represented by different levels of depth changes, this can also help us design an optimal depth codec by increasing or decreasing different levels of depth changes to bring in the smallest VSD. To our best knowledge, the existing methods as aforementioned in subsection I-C can hardly achieve this.


\subsection{Efficiency comparison: running time}
In this subsection, the complexity of these four methods are compared, where the entire frame of VSD prediction is estimated. The average running time of all the 463 frames is shown in TABLE \ref{tab:All}, where the unit is second (s). The running time of the proposed methods listed in TABLE \ref{tab:All} involves two parts. The first part is the running time of the S-VSDs generation. The second part is the running time of the NLM training and testing. The ratio of these two parts is 1000:1 during our test. According to the experimental results, the proposed method is competitive to the state-of-the-art method (e.g., Fang's \cite{fang2013analytical} method) in terms of efficiency. 

Of note, the proposed method is friendly for parallel processing. Each layer can be performed independently during S-VSD calculation, e.g., by a separate thread of the CPU or GPU. Besides, all these anchors except for Fang's \cite{fang2013analytical} method are friendly for parallel processing. After taking the advantages of paralleled design into account, our method outperforms the state-of-the-art in terms of efficiency due to the advantages of our layer-level operations during the S-VSD calculation.

%

\section{Conclusion}
\label{Con}
In this paper, we have proposed an auto-weighted layer representation based view synthesis distortion estimation for 3-D video coding. To achieve this, the level of depth changes and their associated texture degeneration have been used to define the sub view synthesis distortion (S-VSD). After that, a set of theoretical derivations have demonstrated that the VSD can be approximately decomposed into the S-VSDs multiplied by their associated weights. We also have developed a layer-based representation of the S-VSD, where all the pixels with the same level of depth changes are represented with a layer to enable efficient S-VSD calculation. Meanwhile, we have learnt a nonlinear mapping function to better represent the relationship between the VSD and S-VSDs based on our newly built dataset. Experimental results have demonstrated that the proposed method outperforms relevant state-of-the-art VSD estimation methods in both accuracy and efficiency. Besides, unlike existing VSD estimation methods, we propose the first work to relate different levels of depth changes to the VSD. This allows many new applications can be developed for 3-D video coding in our future work, such as optimizing 3-D coding by figuring out the exact contribution of S-VSDs to the VSD, building a more efficient deep codec by increasing and decreasing different levels of depth changes to bring in the smallest VSD, etc.

\ifCLASSOPTIONcaptionsoff
  \newpage
\fi




\end{document}